\documentstyle[preprint,tighten,aps]{revtex}

\begin{document}
\preprint{\begin{minipage}{2in}\begin{flushright}
  UCSD/PTH/97-33 \\ OHSTPY-HEP-T-97-016 \\ hep-lat/9712016 \end{flushright}
  \end{minipage}}
\input epsf
\draft
\title{One-loop matching of lattice and continuum heavy-light
       axial vector currents using NRQCD}
\author{Colin~J.~Morningstar}
\address{Dept.~of Physics, University of California at San Diego,
  La Jolla, California 92093-0319}
\author{J.~Shigemitsu}
\address{Physics Department, The Ohio State University,
  Columbus, OH 43210}
\date{December 15, 1997}
\maketitle

\begin{abstract}
The temporal component of the heavy-light axial vector current is
constructed to one-loop order in perturbation theory and to order $1/M$,
where $M$ is the heavy quark mass, in terms of operators suitable for use
in lattice simulations of $B$ and $D$ mesons.  The $O(a)$-improved clover
action is used for the massless light quark, where $a$ is the lattice
spacing, and propagation of the heavy quark is described by a
nonrelativistic lattice action. 
\end{abstract}
%
%\narrowtext

\section{Introduction}
\label{sec:intro}

Studies of $B$ and $D$ meson decays shed light on many important aspects
of particle physics, such as the extraction of the Cabibbo-Kobayashi-Maskawa
(CKM) matrix elements, CP violation, rare decays, and hints of physics
beyond the standard model.  Such decays involve not only the electroweak
currents, but also strong interaction dynamics in the form of hadronic
matrix elements.  Calculation of these matrix elements is crucial for
high precision tests of the standard model.

A very promising approach to determining hadronic matrix elements from
first principles in QCD is provided by numerical simulations of hadrons
using lattice gauge theory.  Studies of leptonic and semi-leptonic decays
using lattice simulations have already been done using various techniques
to handle heavy quarks on the lattice, such as the static
approach\cite{static}, nonrelativistic QCD (NRQCD)\cite{cornell,lepage},
and a reformulation of conventional Wilson and clover lattice
fermions\cite{fermilab}.  In each approach, the
electroweak currents must be constructed in terms of appropriate operators
comprised from the gluon and quark fields of the lattice theory by
calculating the renormalization factors which relate the hadronic matrix
elements in the lattice-regulated theory to those in the $\overline{MS}$
scheme.  These factors are crucial for both decay constant and semi-leptonic
form factor calculations.  In this paper, we construct the time component
of the heavy-light axial vector current in terms of operators suitable for
simulations\cite{fsubBa,fsubBb} in which the heavy quark is treated using
lattice NRQCD and the light quark propagates according to the
$O(a)$-improved clover action\cite{SWaction}, where $a$ is the lattice
spacing.  The standard Wilson action is used for the gluons.  This
construction is carried out to one-loop order in perturbation
theory and to order $1/M$, where $M$ is the mass of the heavy quark.

The expansion of the heavy-light axial vector current in terms of 
appropriate lattice operators is achieved by matching relevant scattering
amplitudes in perturbation theory.  Consider the heavy-light pseudoscalar
meson decay constant $f_{PS}$ defined in some continuum renormalization
scheme, such as $\overline{MS}$, by,
\begin{equation}
   \langle \, 0 \,| \, A_{\mu} \,|\,PS(p)\, \rangle_{QCD}
   = i\, f_{PS}\,p_{\mu},
\label{deffb}
\end{equation}
where $A_{\mu}$ is the heavy-light axial vector current and
$\vert PS(p)\rangle$ is a pseudoscalar meson state of momentum $p$.  As will
be seen, the axial vector current operator can be written as a sum of
operators $J_{A,{\rm lat}}^{(i)}$ in the lattice-regulated theory.  For
instance, one finds that the temporal component, $\mu = 0$, is a sum of
three operators which contribute through $O(\alpha_s,1/M,a)$, where 
$\alpha_s$ is the QCD coupling:
\begin{equation}  \label{a0}
  A_0 = \sum_{j=0,1,2} C_j(\alpha_s,aM)\   J_{A,{\rm lat}}^{(j)}
 +  O(\alpha_s^2,a^2,1/M^2, \alpha_s a/M).
\end{equation}
Since the role of the dimensionless coefficients $C_j$ is to compensate
for low-energy effects in the current operator from the loss of
short-wavelength QCD modes in the lattice theory, one expects that, due to
asymptotic freedom, they may be computed to a good approximation in
perturbation theory.  They are fixed by requiring that scattering amplitudes
involving these current operators agree in continuum and lattice QCD
to a given order in $\alpha_s$, $a$, and $1/M$.  In a perturbative
determination, a process involving quarks and gluons as asymptotic states
may be used for this matching; one need not consider hadron scatterings
which are much more complicated to evaluate.  Note, however, that the
$C_j$ coefficients are independent of the process chosen.  A one-loop
evaluation of the $C_j$ coefficients then involves the following steps:
(1) select a quark-gluon scattering process induced by the heavy-light 
current and calculate the one-loop amplitude for this process in 
continuum QCD; (2) expand the amplitude in powers of $1/M$; (3) identify
operators in the lattice theory that reproduce the terms in this
expansion; (4) calculate the one-loop mixing matrix of these operators
in the lattice theory; (5) adjust the $C_j$
coefficients to produce a linear combination of lattice current operators
whose one-loop scattering amplitude agrees with that from the
$\overline{MS}$ current to a given order in $1/M$ and $a$.  Note
that the lattice currents are matched directly to the continuum QCD
currents; we do not need to use a two-step matching procedure in which the
lattice currents are first matched to those in a continuum heavy-quark
effective theory\cite{hill,bouchard,bouchard2,borrelli}, which are
then matched to the full QCD currents\cite{golden}.

This paper is organized as follows.  First, in Sec.~\ref{sec:continuum},
we discuss various issues relevant to our matching procedure, such as
the choice of expansion parameter in the perturbation series, the
definition of the heavy quark mass, the regularization of infrared
divergences, and the selection of a renormalization scheme.  Next,
a scattering process appropriate for constructing the heavy-light axial
vector current in the lattice NRQCD-clover theory is chosen.  We then discuss
the continuum QCD calculations, and the necessary current operators in the
lattice regularization scheme are identified.  Sec.~\ref{sec:lattice}
describes the mixing matrix calculation in the lattice theory.  The scattering
amplitudes in the continuum and lattice schemes are matched in
Sec.~\ref{sec:results}, and our main results, the values of the coefficients
$C_j$ for various heavy quark mass values, are presented.  Renormalon
ambiguities are discussed, and subtleties with infrared divergences for
Wilson light quarks are pointed out.  

\section{Continuum Calculation and Operator Identification}
\label{sec:continuum}

Our goal here is the expansion of the temporal component of the
(renormalized) axial vector current in terms of (bare) operators suitable
for use in lattice simulations of $B$ and $D$ mesons.  Note that because
the axial vector current is partially conserved in QCD, its operator
does not actually require renormalization.  This expansion is determined
by matching suitable matrix elements evaluated both in continuum QCD
and in the lattice theory.  Since this expansion is an operator relation,
the expansion coefficients are independent of the external states of the
matrix elements chosen.  Of course, the normalizations of the external
states used in the continuum and lattice matrix elements must match
in order to expose the required operator relation.  Since the external
states in our perturbative calculations are free quark states, this
can be done at tree level by matching the normalizations of the continuum
and lattice Dirac spinors (to the appropriate order in $a$).  The use of
{\em on-shell} wave function renormalization for both the heavy and light
quark fields is then the simplest way to ensure this matching beyond
tree level.

In matching the $1/M$ expansions of heavy-light matrix elements in the
continuum and lattice theories, it is important to establish the relations
between the quark mass parameters (both heavy and light) in each.  Equality
of the lattice and continuum mass parameters can be ensured in perturbation
theory using the pole mass definition.  The pole mass is gauge-invariant and,
within the context of perturbation theory, is a physical observable.
Hence, in our calculations, we use the pole mass for $M$ and also for the
light quark mass $m=0$.  In using the pole mass definition for our quarks,
we compute the continuum and lattice matrix elements using {\em on-shell}
quark mass renormalization.  Note that since the pole mass is not defined
outside of perturbation theory, it is not a useful quantity for lattice
simulations; the bare mass $M_0$ appearing in the lattice action is much
more suitable.  Thus, once we determine the coefficients $C_j(\alpha_s,aM)$
in terms of $M$, we re-express our results in terms of $aM_0$ using the
perturbative relation between $M$ and $M_0$.

We must also know the relationship between the expansion parameters
used in the continuum and lattice perturbative expressions in order to
carry out the operator matching.  For example, we could use
$\alpha_{\overline{MS}}$ in the continuum calculations and a coupling
$\alpha_V$ defined in terms of the momentum-space static quark
potential $V(q)$ in the lattice computations.  The two couplings
are related by $\alpha_V=\alpha_{\overline{MS}}+O(\alpha_s^2)$.  Our
one loop calculations do not make use of the $O(\alpha_s^2)$ correction
in the relation between the lattice and continuum expansion parameters;
hence, as long as we use couplings which match at leading order, we can
simply use $\alpha_s$ to refer to both the lattice and continuum couplings. 
Of course, when numerical values for the coefficients $C_j(\alpha_s,aM)$
are needed, we must specify a renormalization scheme for $\alpha_s$.

The coefficients $C_j(\alpha_s,aM)$ depend on the ultraviolet regulator but
contain no infrared divergences.  However, our choice of on-shell mass
and wave function renormalization will lead to infrared divergences at
intermediate stages in our matrix element calculations.  In order to
demonstrate the cancellation of these divergences in the $C_j$
coefficients, the same infrared regulator must be used in the evaluation
of both the continuum and lattice matrix elements.  Since the triple gluon
vertex plays no role in any of our Feynman diagrams, we introduce a gluon
mass $\lambda$ for this purpose.

Errors in our operator matching are measured in terms of powers of the
lattice spacing $a$, the QCD coupling $\alpha_s$, and the inverse heavy
quark mass $1/M$.  However, NRQCD is a non-renormalizable effective field
theory in which the lattice spacing cannot be taken to zero; rather,
$a=O(1/M)$ must be imposed.  Hence, our error determinations must take
into account the fact that $aM$ is of order unity.  Since we wish to
construct a current operator which is correct through $O(1/M)$, we must
remove all $O(a)$ errors.  We also remove all $O(a/M)$ errors since it is
easy and desirable to do so.  The $O(a,1/M,a/M)$ errors can be eliminated
at tree-level in perturbation theory.  The one-loop calculations are then
used to remove the $O(\alpha_s\, a)$ and $(\alpha_s/M)$ errors.  The
remaining $O(\alpha_s\, a/M)$ errors are much smaller than $O(a)$ and 
$O(\alpha_s\, a)$ and are expected to be comparable to $O(\alpha_s\, a^2)$.

\begin{figure}
\begin{center}
\leavevmode
\epsfxsize=6.0in\epsfbox[50 130 596 782]{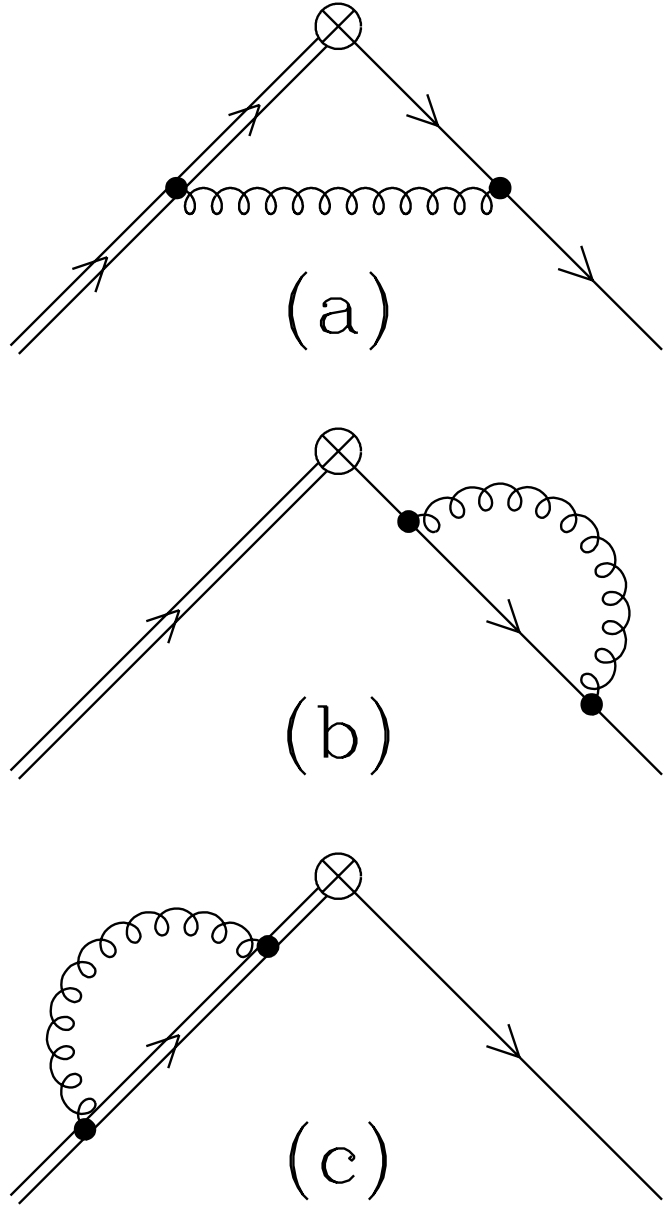}
\end{center}
\caption[fig:feynABC]{
 Feynman diagrams in continuum and lattice perturbation theory which
 contribute at one-loop order to the annihilation by the axial current
 (cross inside a circle) of an incoming heavy quark (double solid line)
 and the creation of an outgoing light quark (single solid line).  The
 exchange of a gluon is denoted by a curly line.
\label{fig:feynABC}}
\end{figure}

A scattering process which is suitable for our study of the heavy-light
axial vector current is depicted in Fig.~\ref{fig:feynABC}.  In this
process, an incoming heavy quark $\vert\,h(p)\,\rangle$ with momentum $p$
scatters off a heavy-light current $A_0$ into an outgoing light quark
$\vert\,q(p^{\prime})\,\rangle$ with momentum $p^{\prime}$.  The 
axial-vector current operator is given by $A_\mu(x) = \bar q(x)\,\hat\gamma_5
\hat\gamma_\mu\,h(x)$ in terms of the light quark field $q(x)$ and the
heavy quark field $h(x)$.  The first
step in our study is the calculation of the amplitude for this
continuum-QCD process to one-loop order in perturbation theory.
Using the on-shell mass and wave function renormalization scheme in Feynman
gauge and expanding in $1/M$ (except for the Dirac spinor of the heavy quark),
one finds 
\begin{eqnarray} \label{fqcdterms}
\langle \, q(p^{\prime}) \,| \, A_0 \, | \,h(p) \, {\rangle}_{QCD}
&= &\quad a_1\,\left[\;\; \bar{u}_q(p^{\prime}) \hat\gamma_5
 \hat\gamma_0 u_h(p)\;\; \right] \quad \;
 + \quad a_2 \,\left[\frac{p_0}{M}\;\bar{u}_q(p^{\prime}) \hat\gamma_5 
 u_h(p)\right] \nonumber \\ 
& & +\;a_3\,\left[ \frac{p \cdot p^{\prime}}{M^2}\;
\bar{u}_q(p^{\prime}) \hat\gamma_5 \hat\gamma_0  u_h(p)\right]  
\; + \; a_4 \,\left[\frac{p_0^{\prime}}{M}\;\bar{u}_q(p^{\prime})
 \hat\gamma_5  u_h(p) \right] \nonumber \\
& & + \;a_5\,\left[ \frac{p \cdot p^{\prime}}{M^2} \frac{p_0}{M} \;
\bar{u}_q(p^{\prime}) \hat\gamma_5  u_h(p)\right] \quad + \quad O(1/M^2),
\end{eqnarray}
with
\begin{eqnarray}
a_1  &=& \ 1\, + \,\frac{\alpha_s}{3 \pi} \left[ \;3\, \ln\frac{M}{\lambda}
\,-\, \frac{11}{4} \right],  \nonumber \\
a_2  &=& \qquad\ \frac{\alpha_s}{3 \pi} \ 2, \nonumber \\
a_3  &=& \qquad\  \frac{\alpha_s}{3 \pi} \left[ \;\;6\, \ln\frac{M}{\lambda}
\,-\,\frac{8\pi}{3} \,\frac{M}{\lambda}\,+ \,\frac{1}{2} \right],\nonumber\\
a_4  &=& \qquad\ \frac{\alpha_s}{3 \pi} \left[- 2\, \ln\frac{M}{\lambda}
\,+\, \frac{1}{2} \right],  \nonumber \\
a_5  &=& \qquad\ \frac{\alpha_s}{3 \pi} \left[ -4\, \ln\frac{M}{\lambda}
\,+\, 5 \right],  
\end{eqnarray}
where $u_h(p)$ and $u_q(p^\prime)$ are the standard spinors for the heavy
and light quarks, respectively, which satisfy the Dirac equation.  The light
quark mass is set equal to zero.  Our conventions for the Dirac
$\hat\gamma$-matrices in Minkowski space are given in the appendix.
Ultraviolet divergences are
regulated using dimensional regularization and fully anti-commuting
$\hat\gamma_5$ matrices are used.  As previously mentioned, we use a gluon
mass $\lambda$ in order to regulate infrared divergences.  The three
diagrams which contribute to this amplitude at one-loop order are shown in
Fig.~\ref{fig:feynABC}.

In lattice NRQCD, the heavy quark is described in terms of a two-component
(in spin space) field $\psi(x)$.  The Dirac field $h(x)$
is related to $\psi$ (and the antiquark
field $\tilde{\psi})$ by a unitary Foldy-Wouthuysen
transformation\cite{FWtrans},
\begin{equation} 
\label{fw}
 h(x) =  U_{FW}^{-1}
  \; \left(\begin{array}{c} \psi(x) \\ \tilde{\psi}(x) \end{array}\right).
\end{equation}
This transformation decouples the upper and lower components of the
Dirac field, thereby separating the quark field from the antiquark field.
To facilitate the identification of lattice NRQCD operators capable of
matching Eq.~\ref{fqcdterms}, we similarly transform the external state
spinor $u_h(p)$ (with normalization $u_h^\dagger u_h=1$) into a
nonrelativistic Pauli spinor:
\begin{equation} \label{fwu}
  u_h(p) =  \left[ 1 - \frac{1}{2 M}(\mbox{\boldmath$\hat\gamma$}
   \cdot {\bf p}) \right] \,u_Q(p) 
\quad + \quad O(1/M^2),
\end{equation}
where
\begin{equation}
 u_Q(p) = \left(\begin{array}{c} U_Q  \\ 0 \end{array} \right),
\end{equation}
and $U_Q$ is a two-component external state spinor depending
only on the spin of the heavy quark.  Using Eq.~\ref{fwu}, the relation
$ \hat\gamma_0 \, u_Q(p) = u_Q(p)$, and the Dirac equation for the light
quark $\bar u_q(p^\prime)p_0^\prime = \bar u_q(p^\prime)
\mbox{\boldmath$\hat\gamma$} \cdot {\bf p^\prime}\hat\gamma_0$,
Eq.~\ref{fqcdterms} may be written
\begin{equation} 
\label{fqcdtermsB}
\langle \, q(p^{\prime}) \,| \, A_0 \, | \,h(p) \, {\rangle}_{QCD}
 =  \eta^A_0\  \hat\Omega_0 + \eta^A_1 \ \hat\Omega_1
 + \eta^A_2\ \hat\Omega_2  \quad + \quad O(\alpha_s^2,1/M^2),
\end{equation}
where $\Gamma = \hat\gamma_5 \hat\gamma_0$, and
\begin{eqnarray}
\hat\Omega_0 & = &  \bar{u}_q(p^{\prime})\ \Gamma\ u_Q(p), \label{omega0}\\
\hat\Omega_1 & = & -  \bar{u}_q(p^{\prime})\ \Gamma
  \frac{\mbox{\boldmath$\hat\gamma$}
   \cdot {\bf p}}{2M}\ u_Q(p), \label{omega1}\\
\hat\Omega_2 & = & - \bar{u}_q(p^{\prime})\ \frac{\mbox{\boldmath$\hat\gamma$}
   \cdot {\bf p^\prime}}{2M}\Gamma\ u_Q(p). \label{omega2}
\end{eqnarray}
The coefficients in Eq.~\ref{fqcdtermsB} may be written
\begin{eqnarray} \label{eta}
\eta^A_0  &=& \;\; (a_1 + a_2)\; = \; 1 + \alpha_s\ \tilde{B}_0,  \nonumber \\
\eta^A_1  &=& \;\;(a_1 - a_2) \;= \; 1 + \alpha_s\ \tilde{B}_1,  \nonumber \\
\eta^A_2  &=&  2\,(a_3 + a_4 + a_5 )\, =  \alpha_s\ \tilde{B}_2,
\end{eqnarray}
where
\begin{eqnarray} \label{Bvalues}
\tilde{B}_0  &=& \frac{1}{3 \pi} \left[ 3\, \ln\frac{M}{\lambda}
- \frac{3}{4} \right],  \nonumber \\
\tilde{B}_1  &=&  \frac{1}{3 \pi} \left[ 3\, \ln\frac{M}{\lambda}
- \frac{19}{4} \right],  \nonumber \\
\tilde{B}_2  &=& \frac{1}{3 \pi} \left[12 - \frac{16 \pi}{3}
 \frac{M}{\lambda} \right].
\end{eqnarray}

Having obtained the $1/M$ expansion of the above scattering amplitude
in continuum QCD, the next step is to identify operators in the lattice
theory which can reproduce the terms in this expansion.
An inspection of Eq.~\ref{fqcdtermsB} suggests immediately that matrix
elements of the following three lattice operators should be considered:
\begin{eqnarray}
 J^{(0)}_{A,{\rm lat}}(x) & = & \bar q(x) \,\Gamma\, Q(x),\label{Jop0}\\
 J^{(1)}_{A,{\rm lat}}(x) & = & \frac{-1}{2M_0} \bar q(x)
    \,\Gamma\,\mbox{\boldmath$\gamma\!\cdot\!\nabla$} \, Q(x),\label{Jop1}\\
 J^{(2)}_{A,{\rm lat}}(x) & = & \frac{1}{2M_0} \bar q(x)
    \,\mbox{\boldmath$\gamma\!\cdot\!\overleftarrow{\nabla}$}
    \,\Gamma\, Q(x), \label{Jop2}
\end{eqnarray}
where $q(x)$ is now the light quark field in the lattice theory,
$M_0$ is the bare heavy quark mass, and $Q(x)$ is related to the heavy
quark field $\psi(x)$ in lattice NRQCD by
\begin{equation}
 Q(x) =  \left( \begin{array}{c} \psi(x) \\ 0 \end{array}\right).
\end{equation}
We use the bare quark mass $M_0$ in the above definitions since
it is the natural mass to use in lattice simulations and because
the pole mass $M$ is not well defined beyond perturbation theory.
The covariant finite difference operators $\nabla_\mu$ and
$\overleftarrow{\nabla}_\mu$ are defined as usual in terms of the
link variables $U_\mu(x)$ which are the parallel transport operators
from sites $x$ to neighboring sites $x\!+\!a_\mu$ in the gauge field.
The definitions of these operators are given below, along with
other derivative operators which will be needed later:
\begin{eqnarray}
 a\nabla_\mu O(x) &=& \frac{1}{2u_0}\left[ U_\mu(x)O(x\!+\!a_\mu)-
    U^\dagger_\mu(x\!-\!a_\mu) O(x\!-\!a_\mu)\right], \\
 O(x)\ a\!\overleftarrow{\nabla}_\mu &=& \frac{1}{2u_0}\left[ O(x\!+\!a_\mu)
  U^\dagger_\mu(x)-O(x\!-\!a_\mu) U_\mu(x\!-\!a_\mu) \right], \\
 a^2\Delta^{(2)} O(x) &=& \sum_{k=1}^{3} \left(
 u_0^{-1}\left[ U_k(x)O(x\!+\!a_k)+U^\dagger_k(x\!-\!a_k)
 O(x\!-\!a_k)\right]-2O(x) \right), \\
 a^2\nabla^{(2)}O(x) &=&  \sum_{\mu=0}^{3} \left(
 u_0^{-1}\left[ U_\mu(x)O(x\!+\!a_\mu)+U^\dagger_\mu(x\!-\!a_\mu)
 O(x\!-\!a_\mu)\right] -2O(x) \right),
\end{eqnarray}
where $O(x)$ is an operator defined at lattice site $x$ with appropriate
color structure, and $u_0$ is the mean link parameter introduced by the
tadpole improvement procedure\cite{viability}.  Note that these lattice
operators are defined in Euclidean space; our Euclidean space conventions
are outlined in the appendix.

\section{Lattice Calculation}
\label{sec:lattice}

In this section, we describe the one-loop calculation in the lattice
theory of the mixing matrix $Z_{ij}$ defined by
\begin{equation} 
\label{latterms}
\langle \, q(p^{\prime}) \,| \, J^{(i)}_{A,{\rm lat}}
  \, | \,h(p) \, {\rangle}_{\rm lat}
 =   \sum_j\ Z_{ij}\ \Omega_j  \ + \ O(\alpha_s^2,1/M^2,a^2,\alpha_s a/M),
\end{equation}
where $\Omega_j$ are the Euclidean-space counterparts of the
$\hat\Omega_j$ defined in Eqs.~\ref{omega0}-\ref{omega2}.  First,
the lattice actions used in these calculations are specified.  The
necessary Feynman diagrams are then presented, and their evaluation
in lattice perturbation theory is outlined.

For the heavy quark, we use the following NRQCD action
density\cite{cornell}:
 \begin{eqnarray} \label{nrqcdact}
  a{\cal L}_{NRQCD} &=&  \psi^\dagger(x)\ \psi(x)\nonumber\\
 &-& \psi^\dagger(x\!+\!a_t)
\left(1 \!-\!\frac{a \delta H}{2}\right)
 \left(1\!-\!\frac{aH_0}{2n}\right)^{n} 
 \frac{U^\dagger_4(x)}{u_0}
 \left(1\!-\!\frac{aH_0}{2n}\right)^{n}
\left(1\!-\!\frac{a\delta H}{2}\right) \psi(x),
 \end{eqnarray}
where 
 \begin{eqnarray}
 H_0 &=& - \frac{\Delta^{(2)}}{2M_0}, \label{Hkin} \\
 \delta H  &=& - c_B \,\frac{g}{2M_0}\,\mbox{\boldmath$\sigma$}
 \cdot{\bf B}\label{deltaH}.
\end{eqnarray}
The positive integer $n$ is introduced to stabilize the highest momentum
modes in the heavy quark propagator\cite{cornell};
the condition, $n > 3/aM_0$, has proven to be a reliable
guide.  Note that the above NRQCD action does not
include the heavy quark mass term. The QCD coupling $g$ is related to
$\alpha_s$ in the usual manner, $\alpha_s=g^2/(4\pi)$, and $\sigma_j$ are the
standard Pauli spin matrices.  At tree level, $c_B = 1$; the one-loop
contribution to $c_B$ is an $O(\alpha_s^2)$ effect in our mixing matrix
calculation and hence can be ignored here.  The chromomagnetic field is
given by $B_j(x)=-\frac{1}{2}\epsilon_{jlm}F_{lm}(x)$,
where the Hermitian and traceless field strength tensor $F_{\mu\nu}(x)$
is defined at the sites of the lattice in terms of clover-leaf operators:
\begin{eqnarray}
 F_{\mu\nu}(x) &=& {\cal F}_{\mu\nu}(x) -
  {1\over 3}{\rm Tr}{\cal F}_{\mu\nu}(x), \nonumber\\
 {\cal F}_{\mu\nu}(x) &=& {-i\over 2a^2g}
\left( \Omega_{\mu\nu}(x)-\Omega^\dagger_{\mu\nu}(x)\right), \nonumber\\
\Omega_{\mu\nu}(x) &=&  {1\over 4u_0^4}\ \sum_{{\lbrace(\alpha,\beta)
   \rbrace}_{\mu\nu}}\!\!U_\alpha(x)U_\beta(x\!+\!a_\alpha)
   U_{-\alpha}(x\!+\!a_\alpha\!+\!a_\beta)U_{-\beta}(x\!+\!a_\beta),
\label{fieldstrength}
\end{eqnarray}
with $\lbrace(\alpha,\beta)\rbrace_{\mu\nu} = \lbrace (\mu,\nu),
(\nu,-\mu), (-\mu,-\nu),(-\nu,\mu)\rbrace$ for $\mu\neq\nu$.

For the light quarks, we use the clover action\cite{SWaction},
\begin{equation}
a{\cal L}_{light} =   \overline{q}\ \mbox{/\hspace{-0.6em}$\nabla$} q 
-  a\,\frac{r}{2} \,\overline{q}\, \nabla^{(2)} q
  + m_0\,\overline{q} q
-iga \frac{r}{4} \sum_{\mu,\nu}\ \overline{q}\, \sigma_{\mu \nu} 
 F_{\mu \nu}\, q,
\label{cloveract}
\end{equation}
where $\mbox{/\hspace{-0.6em}$\nabla$} = \sum_\mu \gamma_{\mu}
 \nabla_{\mu}$, $m_0$ is the bare light quark mass,
$\sigma_{\mu \nu} = \frac{1}{2} \left[\gamma_{\mu} , \gamma_{\nu} 
\right] $, and we set the Wilson parameter $ r = 1$.  The one-loop
correction to the clover coefficient is an $O(\alpha_s^2)$ effect
in our matching calculation and can be neglected.

\begin{figure}
\begin{center}
\leavevmode
\epsfxsize=6.0in\epsfbox[50 60 596 782]{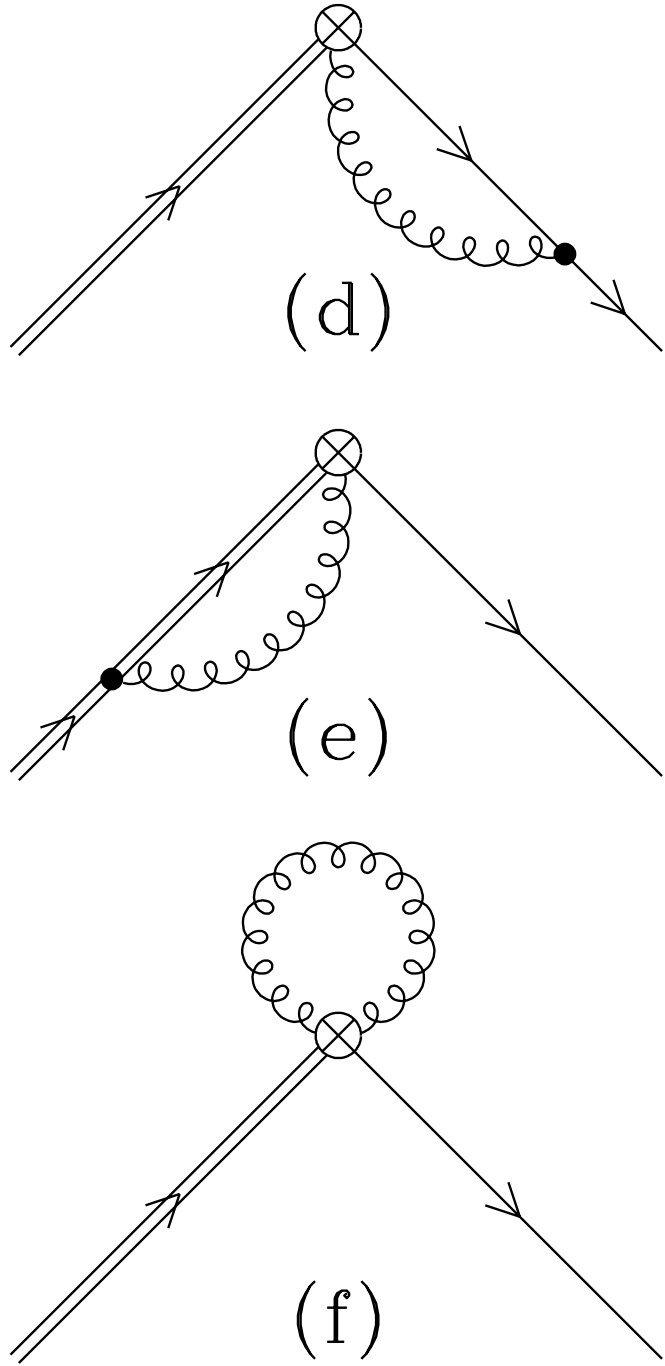}
\end{center}
\caption[fig:feynDEF]{
 Additional vertex correction diagrams which contribute in {\em lattice}
 perturbation theory to the same process as in Fig.~\protect\ref{fig:feynABC}.
\label{fig:feynDEF}}
\end{figure}

\begin{figure}
\begin{center}
\leavevmode
\epsfxsize=6.0in\epsfbox[50 160 596 622]{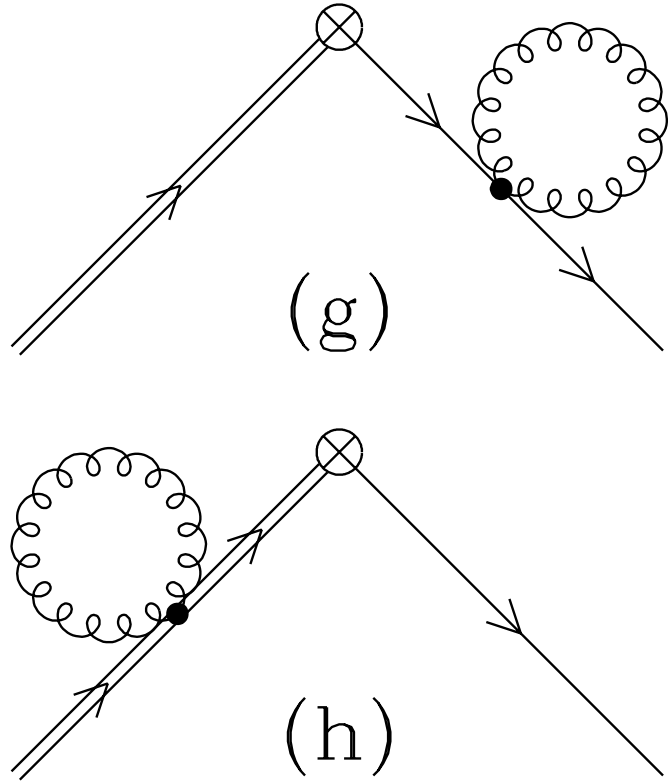}
\end{center}
\caption[fig:feynGH]{
 Additional external leg correction diagrams which contribute in {\em lattice}
 perturbation theory to the same process as in Fig.~\protect\ref{fig:feynABC}.
\label{fig:feynGH}}
\end{figure}

Lattice perturbation theory calculations are much more laborious
than those in continuum perturbation theory.  Not only do more
diagrams contribute to a given process, but the complicated
functional forms of the propagators and vertex functions (presented
in the appendix) necessitate
the use of numerical methods in evaluating the integrals over internal
loop momenta.  For the scattering process considered here, there
are five additional one-loop Feynman diagrams in the lattice theory: three
are vertex corrections, shown in Fig.~\ref{fig:feynDEF}, and two
are extra external leg corrections, shown in Fig.~\ref{fig:feynGH}.
The Feynman rules are determined by expanding the total lattice action
in terms of $g$ using
\begin{equation}
U_\mu(x) \equiv \exp\left[ iagG_\mu\left(x\!+\!\frac{a_\mu}{2}\right)
 \right],
\end{equation}
and $u_0=1-\alpha_s u_0^{(2)}+O(\alpha_s^2)$, then Fourier transforming
into momentum space.  $G_\mu(x)$ is the lattice gluon field defined
at the midpoints of the links connecting neighboring sites.

We calculate the amplitudes corresponding to the Feynman diagrams of
Figs.~\ref{fig:feynABC}-\ref{fig:feynGH} in two very different ways and
verify that the results agree.  In the first method, all spin matrix
manipulations and derivatives with respect to external momenta are
done by hand.  The resulting integrals are also simplified by hand,
including the subtraction of terms to remove infrared divergences.
The remaining infrared-finite integrals are then done numerically
using Monte Carlo and adaptive Gaussian quadrature techniques.

In the second method, nearly all aspects of the calculation are
automated.  First, the propagators and vertex functions are expressed
as functions in C++ whose arguments are the appropriate four momenta
implemented using a class {\tt fourvec}.  Spin algebra is accomplished
using explicit matrix representations;  this is done by defining new class
structures in C++, such as {\tt pauli} and {\tt dirac}, and overloading
all of the necessary arithmetic operators for ease of use.  Derivatives
with respect to external momenta are taken using automatic
differentiation\cite{autodiff}.  This is implemented by defining in C++
a class {\tt Tcomplex} which carries out the multivariate Taylor series
expansions.  A variable of type {\tt Tcomplex} is treated by the end user
just as if it were a regular complex scalar variable since all appropriate
arithmetic operators and mathematical functions are overloaded.  However,
a {\tt Tcomplex} variable is actually an array containing the function
value and all of its derivatives up to some order; an indexing member
function is used to return a specific term in the Taylor series
expansion.  All derivatives are taken automatically using analytical
techniques; the end user never needs to write subroutines for taking these
derivatives. 

The above tools allow one to easily write C++ functions to compute
the integrands for all of our one-loop Feynman diagrams.  The integrals
corresponding to the graphs in Figs.~\ref{fig:feynDEF} and \ref{fig:feynGH}
are easily evaluated using Monte Carlo techniques.  We use the
integration routine {\sc vegas}\cite{vegas}.  Note that, in lattice
perturbation theory, each internal loop momentum $k$ is restricted to the
first Brillouin zone $-\pi< ak_\mu\leq \pi$.  However, due to infrared
divergences, the other integrals cannot be directly calculated.
We evaluate these integrals in a sequence of steps.  First, we subtract
from the lattice integrand $I_{\rm lat}(k)$, where $k$ is the loop
momentum, the analogous integrand from the continuum theory,
$I_{\rm con}(k)$.  The integral of this difference over the first
Brillouin zone is infrared finite and can be evaluated using the Monte
Carlo method.  We then multiply the continuum integrand by a factor $f(k)$
to render the integral insensitive to the shape of the first Brillouin zone.
This factor must also be chosen such that the integral of
$[1\!-\!f(k)]\ I_{\rm con}(k)$ is infrared finite.  For example,
$f(k)=\exp(-\rho k^2)$ where $\rho=\pi/2$ is
often a good choice.  We then use the Monte Carlo method to evaluate
the integral of $[1\!-\!f(k)]\ I_{\rm con}(k)$.  The infrared divergence has
then been isolated in the integral of $f(k)\ I_{\rm con}(k)$.  Since
$f(k)$ is chosen to remove any appreciable sensitivity of the integral
to the edge of the Brillouin zone, we can change to hyperspherical
coordinates and integrate over the interior of an infinitely large sphere
to a good approximation.  This simplifies the calculation and allows us
to easily identify and analytically manipulate the infrared divergent
pieces of the integrand.

Using these two methods, we obtain results for the mixing matrix
$Z_{ij}$ in Eq.~\ref{latterms}, again using the on-shell renormalization
scheme in Feynman gauge.  Note that the nonrelativistic 
external-state spinors $u_Q(p)$ are identical in the lattice and
continuum theories.  For the clover action, the external light-quark
spinors $u_q(p^\prime)$ differ only at $O(a^2)$.  At one-loop, the mixing
matrix elements may be written
\begin{equation} \label{zij}
Z_{ij} = \delta_{ij} + \alpha_s \left[ 
  \frac{1}{2} (\tilde{C}_q + \tilde{C}_Q)\ \delta_{ij}
 + \tilde{C}_m\ \delta_{i1}\delta_{j1} +  \tilde{\zeta}_{ij} \right],
\end{equation}
where $\tilde{C}_q$ and $\tilde{C}_Q$ are the contributions from the light-
and heavy-quark external leg corrections (that is, from wave function
renormalization factors), and $\tilde{\zeta}_{ij}$ denote
the contributions from the vertex corrections.  Our use of an on-shell
renormalization scheme with lattice operators defined in terms of the bare
mass $M_0$ is responsible for the term proportional to $\tilde{C}_m$, where
\begin{equation}
M =  [1 + \alpha_s\, \tilde{C}_m] \, M_0 + O(\alpha_s^2).
\label{massrenorm}
\end{equation}
Note that although the current operators $J_{A,{\rm lat}}^{(i)}$
are defined using $M_0$, the pole mass $M$ must appear in $\Omega_j$.
The factors in Eq.~\ref{zij} may be further decomposed:
\begin{eqnarray}
 \tilde{C}_q &=& C_q + \frac{2}{3\pi}\ln a\lambda +C_q^{\rm TI},\nonumber\\
 \tilde{C}_Q &=& C_Q - \frac{4}{3\pi}\ln a\lambda,\nonumber\\
 \tilde{C}_m &=& C_m + C_m^{\rm TI},\nonumber\\
 \tilde{\zeta}_{ij} &=& \zeta_{ij} + \zeta^{\rm TI}_{ij}
 + \zeta^{\rm IR}_{ij}, \label{zijdecompose}
\end{eqnarray}
where $C_q$, $C_Q$, $C_m$, and $\zeta_{ij}$ are infrared finite and
independent of the tadpole improvement factor $u_0$, and
$\zeta^{\rm IR}_{ij}$ and $\zeta^{\rm TI}_{ij}$
contain the infrared divergences and tadpole improvement contributions,
respectively, from the vertex corrections.  Contributions to $\tilde{C}_q$
and $\tilde{C}_m$ from the tadpole improvement counterterms are
denoted by $C_q^{\rm TI}$ and $C_m^{\rm TI}$, respectively.

\section{Results}
\label{sec:results}

To complete the operator matching, we transform Eq.~\ref{fqcdtermsB} from
Minkowski to Euclidean space, use Eqs.~\ref{latterms} and \ref{zij},
then peel off the dependence on the external states.  This yields the
following operator relation:
\begin{eqnarray} \label{ajalph1}
 A_0 & = & \;\; \left( 1 + \alpha_s \left[ \tilde{B}_0 
 - \frac{1}{2}(\tilde{C}_q+\tilde{C}_Q)
 - \tilde{\zeta}_{00} - \tilde{\zeta}_{10}\right]\right)
 \ J_{A,{\rm lat}}^{(0)}  \nonumber \\
  & & + \left( 1 + \alpha_s \left[ \tilde{B}_1 
 - \frac{1}{2}(\tilde{C}_q+\tilde{C}_Q) -\tilde{C}_m
 -\tilde{\zeta}_{01} - \tilde{\zeta}_{11}\right]\right)
 \ J_{A,{\rm lat}}^{(1)} \nonumber \\
& & + \qquad \;\;  \alpha_s \left[ \tilde{B}_2 -
\tilde{\zeta}_{02} - \tilde{\zeta}_{12}\;\right]
 \ J_{A,{\rm lat}}^{(2)} 
 \quad + \quad O(\alpha_s^2,a^2,1/M^2,\alpha_s a/M).
\end{eqnarray}
As expected, we find that the infrared divergences from the various
terms in the expansion coefficients cancel.  We now have the desired
expansion coefficients $C_j$ of Eq.~\ref{a0} in terms of the pole
mass $M$.  As previously mentioned, the bare mass $M_0$ is a much
more convenient mass parameter.  Our results are easily expressed
in terms of $aM_0$ using Eq.~\ref{massrenorm}, which, in this case,
simply amounts to replacing $M$ by $M_0$:
\begin{eqnarray} \label{ajalph2}
 A_0 & = & \;\; \left( 1 + \alpha_s \left[ B_0 
 - \frac{1}{2}(C_q+C_q^{\rm TI}+ C_Q)
 - \tau_0 \; \right]\right)
 \ J_{A,{\rm lat}}^{(0)}  \nonumber \\
  & & + \left( 1 + \alpha_s \left[ B_1 
 - \frac{1}{2}(C_q+C_q^{\rm TI}+ C_Q) -C_m-C_m^{\rm TI}
 -\tau_1 - \tau_1^{\rm TI}\right]\right)
 \ J_{A,{\rm lat}}^{(1)} \nonumber \\
& & + \qquad \;\;  \alpha_s \left[ B_2 - \tau_2  \;\right]
 \ J_{A,{\rm lat}}^{(2)}
 \quad + \quad O(\alpha_s^2,a^2,1/M^2,\alpha_s a/M),
\end{eqnarray}
where $B_0 = \ln(aM_0)/\pi- 1/(4 \pi)$,
$B_1 = \ln(aM_0)/\pi - 19/(12 \pi)$, $B_2 = 4/\pi$,
$\tau_0 = \zeta_{00} + \zeta_{10}$, $\tau_1 = \zeta_{01} + \zeta_{11}$,
and $\tau_2 = \zeta_{02} + \zeta_{12}$.  The factors arising from
tadpole improvement counterterms are given by
\begin{eqnarray}
 C_q^{\rm TI} &=& -u_0^{(2)},\nonumber\\
 C_m^{\rm TI} &=& -u_0^{(2)} \, \left( 1 
 - \frac{3}{2\, n\, aM_0}\right), \nonumber\\
 \tau_1^{\rm TI} &=& u_0^{(2)}.
\end{eqnarray}
For the usual plaquette definition $u_0=\langle \frac{1}{3}{\rm Tr}
U_\Box\rangle^{1/4}$ in the Wilson gluonic action, $u_0^{(2)}=\pi/3$.
For massless clover quarks, $C_q = 1.030$.  Results for $\tau_0$, $\tau_1$,
$\tau_2$, $C_Q$ and $C_m$ for various values of $aM_0$ are
listed in Table~\ref{tab:one}.  

Explicit calculation reveals that $\tau_2$ behaves as
$-2aM_0\zeta_{disc}$ as $M_0$ becomes large, where $\zeta_{disc}$ is
found numerically to be $\zeta_{disc} = 1.00(1)$.  This factor can
be viewed as arising from a discretization correction
$J_{A,{\rm lat}}^{(disc)}$ to $J^{(0)}_{A,{\rm lat}}$,
\begin{equation}
J_{A,{\rm lat}}^{(disc)} = a \,  \bar q(x)
    \,\mbox{\boldmath$\gamma\!\cdot\!\overleftarrow{\nabla}$}
    \,\Gamma\, Q(x),
\end{equation}
analogous to the $O(a)$ correction $a\ \partial_\mu P$, where $P$
is the pseudoscalar density, to the axial current in light-light quark
systems\cite{alphacollab}.  We can then define an {\em improved} current
operator
\begin{equation}
J_{A,{\rm lat}}^{(0)imp} 
 = J_{A,{\rm lat}}^{(0)} + C_A\, J_{A,{\rm lat}}^{(disc)},
\label{impop}
\end{equation}
where we write 
\begin{equation}
  C_A = \alpha_s \, \left(1+\frac{\zeta_A}{2aM_0}\right).
 \label{CAdef}
\end{equation}
This decomposition is
not unique, and hence, we leave $\zeta_A$ as a free parameter.
Different choices of $\zeta_A$ lead to different $O(\alpha_s^2)$
contributions being included in Eq.~\ref{ajalph2}.
Taking this into consideration, Eq.~\ref{ajalph2} can be written
\begin{eqnarray}
 A_0  &=&  ( 1 + \alpha_s \rho_0) \ J_{A,{\rm lat}}^{(0)imp}
  + ( 1 + \alpha_s \rho_1) \ J_{A,{\rm lat}}^{(1)}
 +  \alpha_s (\rho_2-\zeta_A) \ J_{A,{\rm lat}}^{(2)}\nonumber\\
  &+&  O(\alpha_s^2,a^2,1/M^2,\alpha_s a/M), \label{finalresult}
\end{eqnarray}
so that the $C_j$ coefficients of Eq.~\ref{a0} become
$C_0=1+\alpha_s\rho_0$, $C_1=1+\alpha_s\rho_1$, and 
$C_2=\alpha_s(\rho_2-\zeta_A)$.
This equation is our final result.  Note that at $O(\alpha_s)$,
the $\zeta_A$ dependence in $J_{A,{\rm lat}}^{(0)imp}$ cancels
that in  $C_2 \ J_{A,{\rm lat}}^{(2)}$.
Numerical results for $\rho_0$, $\rho_1$, and $\rho_2$ are given
in Table~\ref{tab:two}.  

In Ref.~\cite{fsubBb}, $B$ meson decay constants were computed
in simulations in which the heavy quark propagation was described
by an NRQCD action different from Eq.~\ref{nrqcdact}; in particular,
higher order corrections were included.  The values of $\rho_0$,
$\rho_1$, and $\rho_2$ for the action used in these simulations are
given in Table~\ref{tab:three}.

In order to use Eq.~\ref{finalresult} in a simulation, a value for
$\alpha_s$ must be specified.  To do this, one must first choose a
renormalization scheme for $\alpha_s$; in the case of a running coupling,
a means of setting the scale must then be devised; lastly, the value of
the coupling at some reference scale must be determined.  A coupling
$\alpha_V(q^\ast)$ defined in terms of the short-distance static potential
with a scale-setting prescription based on the mean value theorem is
advocated in Ref.~\cite{viability}; the value of $\alpha_V(3.402/a)$ can
be obtained from measurements of the average plaquette.  Unfortunately,
we have not computed the $q^\ast$ scales for the $\rho_0$, $\rho_1$, and
$\rho_2$ coefficients.  However, based on findings in Refs.~\cite{hill}
and \cite{morn}, we expect that $q^\ast \sim 2/a$.  An alternative
choice for the expansion parameter is a non-running boosted coupling 
$\alpha_b=6/(4\pi\beta u_0^4)$; the value of this coupling is typically
comparable to $\alpha_V(\pi/a)$.

For the range of $aM_0$ values considered here, our results show no
evidence that perturbation theory is failing.
From Tables~\ref{tab:two} and \ref{tab:three}, one sees that the
one-loop corrections to $C_0$ are very small.  Values for $\rho_0$
are typically near $-0.3$ for all $aM_0$ considered; multiplying
by $\alpha_s\sim 0.2$ yields one-loop corrections to $C_0$ which are
approximately $5\%$ of the tree-level contribution.  For the large
range of $aM_0$ values studied, $\rho_1$ varies between $-0.7$ and $0.5$;
hence, one-loop corrections to $C_1$ are never larger than about $15\%$.
The magnitudes of the values for $\rho_2$ are significantly larger
than those for $\rho_0$ and $\rho_1$.  If we set $\zeta_A=0$, then
$C_A=\alpha_s$ and $C_2=\alpha_s\rho_2$ which, using $\alpha_s\sim 0.2$,
varies between $0.2$ and $0.9$ for the range of $aM_0$ values studied
here.  Since $C_2$ has no tree-level contribution, at least a two-loop
result is needed to check the convergence of its perturbative expansion.

As in all applications of perturbative QCD, nonperturbative
contributions to $C_j$ are possible.  For example,
the perturbative expansions of the $C_j$ coefficients contain renormalon
ambiguities arising from a slight mismatch between the infrared physics
of the lattice theory and that of continuum QCD.  This mismatch is caused
by discretization artifacts and the truncation of the $1/M$ expansion.
In our calculation, contributions coming from momenta small relative to a
given infrared scale $q_0\ll 1/a,M$ are all suppressed by $a^2q_0^2$ or
$q_0^2/M^2$ since the infrared structure of our lattice theory is designed
to agree with continuum QCD through~$O(a,1/M)$.  This infrared suppression
is not true diagram-by-diagram (some diagrams are even infrared divergent),
but the infrared contribution is suppressed by $a^2$ or $1/M^2$ when all
diagrams are summed.  For example, in some analyses\cite{sac}, coefficients
like $C_0$ are separated into the contribution $c_0$ from the leading operator
(here $J_{A,{\rm lat}}^{(0)}$) and contributions $\tilde{c}_0$ from mixings
with higher-dimension operators (here $J_{A,{\rm lat}}^{(1)}$ and 
$J_{A,{\rm lat}}^{(2)}$).  In our analysis, such a separation is artificial
and would be a mistake since it induces $O(a,1/M)$ ambiguities in the separate
pieces $c_0$ and $\tilde{c}_0$. These ambiguities cancel when the pieces
are recombined, leaving contributions to $C_0$ suppressed by $a^2q_0^2$ or
$q_0^2/M^2$.  Thus, any renormalon ambiguities are suppressed by the same
factors and are at worst comparable to the other truncation errors in the
analysis.

Our calculation differs from conventional lattice calculations in
that the $a^2$~truncation errors, perturbative and nonperturbative, cannot
be made arbitrarily small by reducing the lattice spacing\cite{cornell}. 
This is because NRQCD has nonrenormalizable interactions whose couplings
do not vanish as $a\to 0$. The perturbative expansions for $C_j$ have
power-law terms of the form $\alpha_s/(aM)^n$ which ruin the
convergence of perturbation theory if $a$ is taken to zero.
In practice, this problem is avoided by ensuring that $1/(aM)$ is not
large.  Our inability to take $a\to0$ in NRQCD fundamentally limits the
precision of our $O(a,1/M)$~accurate formalism. If improved precision is
needed, we must reduce the truncation errors by using more accurate
discretizations of the lattice currents and action. In this way
truncation errors, renormalon ambiguities, etc.\ are pushed off to
$O(a^3,1/M^2)$ or higher.  

This approach differs dramatically from that used in the static-quark
expansion; in the latter approach, problems with power divergences arise
when the limit $a\to0$ is attempted in a brute-force manner.   The
role of the power-law terms in the $C_j$ coefficients is to cancel
unphysical power-divergent contributions in the lattice NRQCD current
matrix elements, thereby ensuring the correct matching to full continuum
QCD through the relevant order in $\alpha_s$ and $1/M$.  If the
unphysical terms behaving as $\alpha_s/(aM)^n$ in the $C_j$ coefficients
and the current matrix elements become comparable to or larger than the
physical terms of $O(\Lambda_{QCD}/M)$, this cancellation becomes a very
delicate issue and the need for nonperturbative subtraction methods may
arise.  Such is the case in the static-quark expansion; taking the
brute-force limit $a\to0$ obscures the physically relevant contributions
by amplifying the power-divergent unphysical terms.  Here, this problem
is circumvented by keeping the lattice spacing large enough that $1/(aM)$
never becomes large.  The success of low-order perturbation theory in
cancelling the unphysical terms depends, of course, on the relative size
of the physical contributions.  Recent numerical simulations\cite{fsubBb,Hein}
find no evidence of large power-law contamination in the current
matrix elements for the range of $aM_0$ appropriate for $b$-quark
physics, suggesting that power-divergent nonperturbative and higher-order
perturbative contributions to the matching coefficients are also not large.
This finding applies to the temporal component of the heavy-light axial
vector current.  For other operators, the situation could be different
and each case must be examined separately.  

The results in Table~\ref{tab:two} are suitable for values of $aM_0$
appropriate for current simulations of heavy-light systems using NRQCD.
However, terms proportional to $\ln(aM_0)$ cause the $O(\alpha_s)$
contributions to $C_0$ and $C_1$ to become large as $aM_0$ becomes
large.  In such cases, the renormalization group should be used to
improve upon the estimates from one-loop perturbation theory.
Since the left-hand side of Eq.~\ref{a0} is independent of the lattice
spacing, it follows that, neglecting $O(a^2)$ terms,
\begin{equation}
 a\frac{d}{da}\left[ \sum_k C_k(\alpha_s,aM_0)
 \ \langle J^{(k)}_{A,{\rm lat}}\rangle\right] = 0.
\end{equation}
Using Eq.~\ref{latterms} and $d\,\Omega_j/da=0$, a renormalization group
equation for the $C_j$ coefficients, collected into a vector $\vec{C}$,
can be obtained which describes the change in $C_j$ as the lattice spacing
is varied:
\begin{equation}
 \left(a\frac{d}{da}+\gamma^{\rm tr}\right) \vec{C} = 0,
\label{rg1}
\end{equation}
where the anomalous dimension matrix is given by
\begin{equation}
\gamma_{ij}(\alpha_s,aM_0) = \sum_k
 \left(a\frac{d}{da}Z_{ik}\right) Z^{-1}_{kj}.
\end{equation}

In the limit of large $aM_0$, we find 
$\gamma={\rm diag}(-\alpha_s/\pi,-\alpha_s/\pi,-\alpha_s/\pi)$ and
Eq.~\ref{rg1} can be easily solved.  First, express the $C_j$
coefficients as a function of $\alpha_s(a)$ and $aM$ instead of
$\alpha_s(a)$ and $aM_0(a)$ since $M$ is fixed and does not run
with $a$.  If the lattice theory is renormalized in such a way that the
renormalization group $\beta$-function $\beta_{\rm RG}
=-a\,d\alpha_s(a^{-1})\;/da$ is independent of $M_0$ and $m_0$, then
\begin{equation}
\beta_{\rm RG}(\alpha_s) = -2\beta_0\,\alpha_s^2 -2\beta_1\,\alpha_s^3 
 + O(\alpha_s^4),
\end{equation}
where $\beta_0=(11-\frac{2}{3}N_f)/(4\pi)$ 
and $\beta_1=(102-\frac{38}{3}N_f)/(16\pi^2)$, for $N_f$ light quark
flavors.   For large, fixed $M$, Eq.~\ref{rg1} then tells us that the
change in $C_0$ and $C_1$ in going from an initial lattice spacing $a_1$
to a smaller lattice spacing $a_2$ is given, in leading logarithmic
approximation, by
\begin{equation}
  \frac{C_i(a_2M)}{C_i(a_1M)} =
 \left[ \frac{\alpha_s(1/a_2)}{\alpha_s(1/a_1)}\right]^{1/(2\beta_0\pi)},
\label{rg3}
\end{equation}
for $i=0,1$, where $\alpha_s(\mu)$ is the familiar QCD running coupling;
its two-loop form is
\begin{equation}
  \alpha_s(\mu)=\left[\beta_0\ln(\mu^2/\Lambda^2)+\frac{\beta_1}{\beta_0}
 \ln\ln(\mu^2/\Lambda^2)\right]^{-1}.
\end{equation}
Let $a_2M=aM_0^\prime$ and $a_1 M=aM_0$, then
\begin{eqnarray}
\frac{\alpha_s(1/a_2)}{\alpha_s(1/a_1)} &=&
1-\alpha_s\beta_0\ln\left(\frac{a_1^2}{a_2^2}\right) 
+ O(\alpha_s^2),\nonumber\\
&=& 1-\alpha_s\beta_0\ln\left(\frac{M_0^2}{M_0^{\prime 2}}\right) 
+ O(\alpha_s^2),
 \nonumber\\
&=& \frac{\alpha_s(M_0)}{\alpha_s(M_0^\prime)} + O(\alpha_s^2).
\end{eqnarray}
Hence, to the order at which we are working, Eq.~\ref{rg3} is equivalent to
\begin{equation}
  \frac{C_i(aM_0^\prime)}{C_i(aM_0)} =
 \left[ \frac{\alpha_s(M_0^\prime)}{\alpha_s(M_0)}\right]^{-1/(2\beta_0\pi)},
\label{rg4}
\end{equation}
for large $aM_0$ and $aM_0^\prime$.  When $aM_0$ is not large, the
dependence of $\gamma_{ij}$ on $aM_0$ cannot be neglected and
Eq.~\ref{rg4} must be suitably modified.

As an aside, we mention the following fact.  If one wishes to
use the Wilson action instead of the clover action for the light
quark, one cannot include the discretization correction term 
$J_{A,{\rm lat}}^{(disc)}$.  If one includes $J_{A,{\rm lat}}^{(disc)}$
while using Wilson light quarks, one finds an uncancelled logarithmic
infrared divergence upon attempting to match the continuum and lattice
scattering amplitudes.  This divergence is removed by including contributions
from the $O(a)$-correction term in the clover action.

\section{Summary}
\label{sec:summary}

In this paper, the temporal component of the heavy-light axial vector
current $A_\mu$ was expanded in terms of lattice operators suitable
for use in simulations of $B$ and $D$ mesons.  The expansion
was carried out to $O(1/M)$ by matching relevant scattering amplitudes
to one-loop order in perturbation theory.  The (massless) light quark
was described in the lattice theory using the $O(a)$-improved clover
action of Eq.~\ref{cloveract}, and the NRQCD action of Eq.~\ref{nrqcdact}
was used to treat the heavy quark.  The standard Wilson action was
used for the lattice gluons.  The expansion of the heavy-light current
was found to be
\begin{equation}
 A_0  =  C_0 \ J_{A,{\rm lat}}^{(0)imp}
  + C_1 \ J_{A,{\rm lat}}^{(1)}
 +  C_2 \ J_{A,{\rm lat}}^{(2)}  +  O(\alpha_s^2,a^2,1/M^2,\alpha_s a/M),
\label{fres}
\end{equation}
where the lattice operators are defined in Eqs.~\ref{Jop0}, \ref{Jop1},
\ref{Jop2}, and \ref{impop}, and $C_0=1+\alpha_s\rho_0$, 
$C_1=1+\alpha_s\rho_1$, 
$C_2=\alpha_s(\rho_2-\zeta_A)$ and $C_A=\alpha_s[1+\zeta_A/(2aM_0)]$.
Values for $\rho_0$, $\rho_1$, and $\rho_2$ are listed in
Table~\ref{tab:two} for various bare heavy quark masses.  $\zeta_A$
remains as a free parameter; different choices for the value of
$\zeta_A$ lead to different $O(\alpha_s^2)$ contributions being
included in Eq.~\ref{fres}.  The one-loop corrections for $C_0$ and $C_1$
were shown to be small relative to the tree-level contributions for the
range of $aM_0$ studied; no evidence of a breakdown in perturbation
theory was found.  Since $C_2$ has no tree-level contribution, at least
a two-loop calculation would be needed to check the behavior of its
perturbative expansion.  Renormalon ambiguities were argued to be
at worst of the same order as our other systematic errors. 

Our results have already been used in $B$-meson simulations\cite{fsubBa}
measuring $f_B$.  More recently, $B$ meson decay constants were
computed\cite{fsubBb} in simulations using an NRQCD action with
higher-order interactions not included in Eq.~\ref{nrqcdact}; the
values of $\rho_0$, $\rho_1$, and $\rho_2$ for the action used in
these simulations are given in Table~\ref{tab:three}.  In the future,
we plan to apply the methods described in this paper to the expansion
of other currents, such as the vector current, in terms of appropriate
lattice operators.

\section{Acknowledgments}
We would like to thank Peter Lepage, Arifa Ali Khan, Tanmoy
Bhattacharya, and Geoff Bodwin for helpful discussions.  We would
also like to thank Christine Davies for checking some of our
continuum QCD calculations.
This work was supported by the U.S.~DOE, Grants No.~DE-FG03-90ER40546
and DE-FG02-91ER40690, and by NATO grant CRG 941259.

\appendix

\section{Feynman rules}
\label{sec:feynrules}

The NRQCD and light quark actions are given in Eqs.~\ref{nrqcdact}
and \ref{cloveract} in Sec.~\ref{sec:lattice}.  The Feynman rules of
perturbation theory can be derived from these actions using, for instance,
the methods of Ref.~\cite{cjfrules}.  Most of the rules relevant for the
calculations of this article are collected in this appendix.  Various
conventions used in our computations are also outlined.  To simplify
notation, we set the lattice spacing $a=1$ in this appendix.

Minkowski-space quantities are indicated either by a caret, such as
$\hat{\gamma}$, or by a subscript or superscript $(M)$, such as $x_j^{(M)}$.
The metric tensor in Minkowski space is taken to be 
$g_{\mu\nu} = {\rm diag}(1,-1,-1,-1)$ and the
Dirac matrices satisfy $\{\hat\gamma_\mu , \hat\gamma_\nu \} 
= 2 \, g_{\mu \nu}$.  We use the Dirac-Pauli representation:
\begin{eqnarray*}
 \hat\gamma^0 = \hat\gamma_0 &=& 
 \left(\begin{array}{cc} 
        I &\; 0\\
        0 &  -I
  \end{array}  \right),\\
\hat\gamma^j = - \hat\gamma_j &=&
 \left(\begin{array}{cc} 
     \;   0 & \sigma_j \\
  -\sigma_j &  0
 \end{array}  \right),
\end{eqnarray*}
where $\sigma_j$ are the standard Pauli spin matrices.  Also, we define
$\hat\gamma_5 = i \hat\gamma^0 \hat\gamma^1 \hat\gamma^2 \hat\gamma^3.$

Euclidean-space four-vectors are defined in terms of Minkowski-space
four-vectors using $ x_0 = x^0 = i\,x^0_{(M)}$ and $x_j = x^j = 
x^j_{(M)} = - x_j^{(M)}$, for $j=1,2,3$.  For the derivative operator,
$\partial_0 = \partial^0 = -i \partial_0^{(M)}$ and
$\partial_j = \partial^j = \partial_j^{(M)} = - \partial^j_{(M)}$.
Note that the gauge field $G^\mu_{(M)} = ( \phi_{(M)}, \vec{G}_{(M)} )$
Wick-rotates into Euclidean space as a covariant vector, as does the
gauge-covariant derivative $D^\mu_{(M)}$.  The Euclidean-space Dirac
matrices satisfy $ \{\gamma_\mu , \gamma_\nu \} = 2 \, \delta_{\mu \nu}$
and are related to their Minkowski-space counterparts by
$ \gamma_0 = \gamma^0 = \hat{\gamma}^0$,
$\gamma_j = \gamma^j = -i\,\hat{\gamma}^j = i \,\hat{\gamma}_j$, and
$\gamma_5 =  \gamma_0 \gamma_1 \gamma_2 \gamma_3 = \hat{\gamma}_5$.
Also, $\sigma_{\nu \mu} \equiv \frac{1}{2} \, [\gamma_\nu , \gamma_\mu ]$.
To be consistent with our conventions for Euclidean-space 
$\gamma$ matrices, we define Euclidean-space quark-bilinear axial and 
vector currents by $ A_0 = A_0^{(M)}$, $ A^j = -i \, A^j_{(M)} $,
$V_0 =  V_0^{(M)}$, and $ V^j =  -i \, V^j_{(M)} $.  The chromoelectric
and chromomagnetic fields are defined in terms of the field strength
tensor by $ E_j = F_{0j} = -i\,F_{0j}^{(M)} = i \, E_j^{(M)}$ and
$ B_j = - \frac{1}{2}\varepsilon_{lmj} F_{lm}  
= - \frac{1}{2}\varepsilon_{lmj} F_{lm}^{(M)} = B_j^{(M)} $, where
$\varepsilon_{ijk}$ is the fully antisymmetric Levi-Civita tensor and
the field strength tensor is given on the lattice
by Eq.~\ref{fieldstrength}.

The heavy quark propagator is diagonal in both spin and color, and
is given in momentum space by
\begin{eqnarray*}
 \tilde{G}_Q(k)  &=& \left(1 - e^{-ik_0} \,F^{2n}(k) \right)^{-1},\\
  F(k) &=&  1 - \frac{1}{nM} \sum_{j=1}^3 
  \sin^2({\textstyle\frac{1}{2}}k_j).
\end{eqnarray*}
The light quark propagator is diagonal in color:
\[
\tilde{G}_q(k) = \left( i \sum_\mu \gamma_\mu \sin\,k_\mu + 
2 r \sum_\mu \sin^2({\textstyle\frac{1}{2}}k_\mu)
 + m \right) ^{-1}.
\]
In this paper, we work only with the $m = 0$ and $r=1$ case.
The gluon propagator is diagonal both in color and the Lorentz indices,
and is given in Feynman gauge by
\[
 \tilde{G}_G(k) = \left( 4 \sum_\mu \sin^2({\textstyle\frac{1}{2}}
 k_\mu) + \lambda^2 \right)^{-1}.
\]

We now list the vertex factors associated with the interaction of
a single gluon with a heavy-quark line.  Let $k^{\prime}$ be the
outgoing heavy-quark momentum, $k$ be the incoming quark momentum,
and $\mu$ be the polarization index of the emitted gluon.
These vertex factor all have a color
factor of $T^a_{bc}$, where $a$ is the color index of the gluon,
and $b$ and $c$ are the color indices of the outgoing and incoming
quarks, respectively.  For $\mu=0$, the vertex factor is
\[
-ig e^{-i(k^{\prime} + k)_0/2} \; F^n(k)\; F^n(k^{\prime}).
\]
For $\mu = j=1,2,3$, the vertex factor from the ${\bf p}^2/2M$ term is
\begin{eqnarray*}
 &-g \displaystyle{\;\frac{1}{2nM}\;} \sin[{\textstyle\frac{1}{2}}
 (k^{\prime} + k)_j] \,
\left[ e^{-ik^{\prime}_0}  F^n(k^{\prime})+ e^{-ik_0} F^n(k)\right]
 \;S_{n}(k^{\prime},k),& \\
 &S_{n}(k^{\prime},k) = 
 \sum_{l=0}^{n-1} F^{l}(k^{\prime})\, F^{n-l-1}(k).&
\end{eqnarray*}
For $\mu  = j$ from the $\mbox{\boldmath$\sigma$} \cdot {\bf B}$ term,
the vertex factor is given by
\[ g \;\sum_{r,s}\frac{1}{4M}\;(\gamma_5 \gamma_0 \gamma_r)\;\epsilon_{rsj} 
\;\sin(k^{\prime} - k)_s \; \cos[{\textstyle\frac{1}{2}}
 (k^{\prime} - k)_j] \;
\left[ e^{-ik^{\prime}_0}  F^{2n}(k^{\prime})+ e^{-ik_0} F^{2n}(k)\right].
\]

The single-gluon vertex factor for the light quark (clover action) is
\[
 -g \left\{ i \gamma_\mu  \cos[{\textstyle\frac{1}{2}}(k^{\prime} 
+ k)_\mu]  +  r  \sin[{\textstyle\frac{1}{2}}
 (k^{\prime} + k)_\mu] 
 +  \frac{r}{2}  \sum_\nu \sigma_{\nu \mu}
 \sin(k^{\prime} - k)_\nu  \cos[{\textstyle\frac{1}{2}}
(k^{\prime} - k)_\mu]\right\},
\]
for all gluon polarizations $\mu$, where $k^\prime\ (k)$ is the
outgoing (incoming) momentum of the light quark.

Next, the vertex factors associated with the interaction of
two gluons with a heavy-quark line are given.  Let $k^{\prime}$ be the
outgoing heavy-quark momentum, $k$ be the incoming quark momentum,
and $q_1$, $q_2$ be the outgoing momenta of the emitted gluons having
polarization indices $\mu_1$, $\mu_2$, respectively.  Only the factors
for $\mu_1 = \mu_2$ (and $q_1 + q_2 = 0$) are given, which is all 
that is required for the one-loop tadpole graphs.  The color factor
for these cases is $(T^a\;T^{a^\prime})_{bc}$, where $a$, $a^\prime$
are the color indices of the gluons, and $b$ and $c$ are the color
indices of the outgoing and incoming quarks, respectively. 
For $\mu_1 = \mu_2 = 0$, the vertex factor is
\[
- \frac{g^2}{2} \; 
e^{-i(k^{\prime} + k)_0/2}\;F^n(k^{\prime}) \; F^n(k).
\]
For $\mu_1 = \mu_2 = j$, the vertex factor from the ${\bf p}^2/2M$
term in the NRQCD action is
\begin{eqnarray*}
&g^2 \,\left[ e^{-ik^{\prime}_0}  F^n(k^{\prime})
 + e^{-ik_0} F^n(k)\right]
\;\biggl(\displaystyle{\frac{-1}{4nM}}\; \cos[{\textstyle\frac{1}{2}}
 (k^{\prime}\!+\!k)_j] \; 
S_{n}(k^{\prime},k) &\\
&+  {\displaystyle \frac{1}{(2nM)^2}}\;\sin[{\textstyle\frac{1}{2}}
 (k^{\prime}\!+\!k\!-\!q_2)_j] \;
\sin[{\textstyle\frac{1}{2}}(k^{\prime}\!+\!k\!+\!q_1)_j] \; 
S_{g,n}(k^{\prime},k^{\prime}+q_1,k)\biggr) &\\
&+ g^2 \;  {\displaystyle\frac{1}{(2nM)^2}}
 \;\sin[{\textstyle\frac{1}{2}}(k^{\prime}\!+\!k\!-\!q_2)_j] \;
\sin[{\textstyle\frac{1}{2}}(k^{\prime}\!+\!k\!+\!q_1)_j]
 \; S_{n}(k^{\prime},k-q_2) \;
S_{n}(k^{\prime}+q_1 , k) \; e^{-i(k^{\prime} + q_1)_0} ,
\end{eqnarray*}
where
 \[ 
 S_{g,n}(k^{\prime}, k^{\prime}+q_1, k) = 
\sum_{l\geq 0}^{n-2} F^{l}(k^{\prime})\;S_{n-l-1}(k^{\prime}+q_1,k).
\]
The vertex factor for $\mu_1 = \mu_2 = j$ from the
$\mbox{\boldmath$\sigma$} \cdot {\bf B}$ term is
\begin{eqnarray*}
& g^2\;{\displaystyle\frac{1}{(4M)^2}}\;\sum_{r,s,r^\prime,s^\prime}
 \; e^{-i(k^{\prime}+q_1)_0} \;F^{2n}(k^{\prime}+q_1) 
\;(\gamma_5\gamma_0\gamma_r)\;(\gamma_5\gamma_0\gamma_{r^\prime}) 
\; \epsilon_{rsj}\; \epsilon_{r^\prime s^\prime j} &\\
& \times\; \sin(k^{\prime}-k+q_2)_s
\;\sin(k^{\prime}-k+q_1)_{s^\prime} \; \cos[{\textstyle\frac{1}{2}}
(k^{\prime}-k+q_2)_j] 
\; \cos[{\textstyle\frac{1}{2}}(k^{\prime}-k+q_1)_j] & \\
&+ \mbox{\rm terms that vanish for $q_1+q_2 = 0$}.
\end{eqnarray*}

The heavy-light currents, $J_{A,{\rm lat}}^{(i)}$, are not part of the
action and we list their Feynman rules separately.  Again, let
$k^{\prime}$ and $k$ be the outgoing and incoming quark momenta,
respectively, and let $q_i$ and $\mu_i$ denote the momenta and
polarizations, respectively, of the emitted gluons.  Also, let
$q_{ext}$ be the momentum carried off by the external heavy-light current,
where $ k = k^{\prime} + \sum_iq_i + q_{ext}$.  We use the notation
$[O]^{(n)}$ to indicate the vertex factor for $n$-gluon emission from
operator $O$.  At tree level,
\begin{eqnarray*}
 \left[ J^{(1)}_{A,{\rm lat}} \right]^{(0)} &=& - \frac{i}{2M} \; \sum_j 
\sin[{\textstyle\frac{1}{2}}(k^{\prime} + k + q_{ext})_j]
 \; \left(\Gamma \, \gamma_j \right),\\
 \left[ J^{(2)}_{A,{\rm lat}} \right]^{(0)} &=& - \frac{i}{2M} \; \sum_j 
\sin[{\textstyle\frac{1}{2}}(k^{\prime} + k - q_{ext})_j]
 \; \left( \gamma_j \, \Gamma \right).
\end{eqnarray*}
For one gluon emission ($\mu = j$, color factor = $T^a_{bc}$):
\begin{eqnarray*}
\left[ J^{(1)}_{A,{\rm lat}} \right]^{(1)} &=& - \;g \;\frac{i}{2M} \; 
\cos[{\textstyle\frac{1}{2}}(k^{\prime} + k + q_{ext})_j] \; 
\left(\Gamma \, \gamma_j \right),\\
 \left[ J^{(2)}_{A,{\rm lat}} \right]^{(1)} &=& - \;g \;\frac{i}{2M} \; 
\cos[{\textstyle\frac{1}{2}}(k^{\prime} + k - q_{ext})_j] \;
 \left( \gamma_j \, \Gamma \right).
\end{eqnarray*}
For two gluon emission ($\mu_1 = \mu_2 = j$, color factor =
$(T^a T^{a^{\prime}})_{bc}$):
\begin{eqnarray*}
\left[ J^{(1)}_{A,{\rm lat}} \right]^{(2)} &=& g^2 \; \frac{i}{4M} \; 
\sin[{\textstyle\frac{1}{2}}(k^{\prime} + k + q_{ext})_j]
 \; \left(\Gamma \, \gamma_j \right),\\
\left[ J^{(2)}_{A,{\rm lat}} \right]^{(2)} &=& g^2 \; \frac{i}{4M} \; 
\sin[{\textstyle\frac{1}{2}}(k^{\prime} + k - q_{ext})_j]
 \; \left( \gamma_j \, \Gamma \right).
\end{eqnarray*}

%%%%%%%%%%%%%%%%%%%%%%%%%%%%%%%%%%%%%%%%%%%%%%%%%%%%%%%%%%%%%%%%%%%%%
%%
%%                             REFERENCES
%%
%%%%%%%%%%%%%%%%%%%%%%%%%%%%%%%%%%%%%%%%%%%%%%%%%%%%%%%%%%%%%%%%%%%%%

%%%%%%%%%%%%%%%%%%%%%%%%%%%%%%%%%%%%%%%%%%%%%%%%%%%%%%%%%%%%%%%%%%%%%
%%
%%                               FIGURES
%%
%%%%%%%%%%%%%%%%%%%%%%%%%%%%%%%%%%%%%%%%%%%%%%%%%%%%%%%%%%%%%%%%%%%%%

\newpage

\newpage

\newpage

%%%%%%%%%%%%%%%%%%%%%%%%%%%%%%%%%%%%%%%%%%%%%%%%%%%%%%%%%%%%%%%%%%%%%
%%
%%                                TABLES
%%
%%%%%%%%%%%%%%%%%%%%%%%%%%%%%%%%%%%%%%%%%%%%%%%%%%%%%%%%%%%%%%%%%%%%%

\begin{table}
\caption[tabone]{Values of the coefficients $\tau_0$, $\tau_1$,
  $\tau_2$, $C_Q$, and $C_m$ appearing in Eq.~\protect\ref{ajalph2}
  for various values of the bare heavy-quark mass $aM_0$ and NRQCD
  stability parameter $n$.  Uncertainties in the determinations of these
  parameters due to the use of Monte Carlo integration are included.
\label{tab:one}}
\begin{center}
\begin{tabular}{rrrrrrr}
\multicolumn{1}{c}{$aM_0$} &
\multicolumn{1}{c}{$n$} &
\multicolumn{1}{c}{$\tau_0$} &
\multicolumn{1}{c}{$\tau_1$} &
\multicolumn{1}{c}{$\tau_2$} &
\multicolumn{1}{c}{$C_Q$} &
\multicolumn{1}{c}{$C_m$} \\ \hline
10.0 & 1 & 0.8232(1) &$-1.242(4) $ & $-14.74(2)$\hspace{0.5em} 
   & $ 0.2719(8)$ & 0.872(8)\\
 7.0 & 1 & 0.7929(1) &$-1.266(1) $ & $-8.82(2) $\hspace{0.5em}
   & $ 0.1847(6)$ & 0.862(6)\\
 4.0 & 1 & 0.7290(1) &$-1.283(2) $ & $-3.202(8)$ & $-0.0287(6)$ & 0.883(6)\\
 4.0 & 2 & 0.7397(1) &$-1.298(2) $ & $-3.476(8)$ & $-0.0030(6)$ & 1.087(6)\\
 3.5 & 2 & 0.7239(1) &$-1.307(2) $ & $-2.618(8)$ & $-0.0651(6)$ & 1.120(6)\\
 3.0 & 2 & 0.7052(1) &$-1.315(2) $ & $-1.790(8)$ & $-0.1463(6)$ & 1.142(6)\\
 2.7 & 2 & 0.6923(1) &$-1.324(2) $ & $-1.314(4)$ & $-0.2077(6)$ & 1.172(6)\\
 2.5 & 2 & 0.6829(1) &$-1.332(2) $ & $-1.010(4)$ & $-0.2562(6)$ & 1.195(6)\\
 2.0 & 2 & 0.6576(1) &$-1.352(1) $ & $-0.304(3)$ & $-0.4156(6)$ & 1.235(4)\\
 1.7 & 2 & 0.6422(1) &$-1.371(2) $ & $ 0.081(3)$ & $-0.5503(8)$ & 1.277(4)\\
 1.6 & 2 & 0.6377(1) &$-1.377(1) $ & $ 0.202(3)$ & $-0.6038(8)$ & 1.293(6)\\
 1.2 & 3 & 0.6307(1) &$-1.424(1) $ & $ 0.538(2)$ & $-0.868(1)\hspace{0.5em} $
   & 1.557(6)\\
 1.0 & 4 & 0.6365(1) &$-1.457(1) $ & $ 0.683(2)$ & $-1.074(1)\hspace{0.5em} $
   & 1.733(6)\\
 0.8 & 5 & 0.6617(1) &$-1.502(1) $ & $ 0.818(1)$ & $-1.394(1)\hspace{0.5em} $
   & 1.934(6)
\end{tabular}
\end{center}
\end{table}

\begin{table}
\caption[tabtwo]{Values for the coefficients $\rho_0$, $\rho_1$,
 and $\rho_2$ appearing in Eq.~\protect\ref{finalresult}.
\label{tab:two}}
\begin{center}
\begin{tabular}{rrrrl}
\multicolumn{1}{c}{$aM_0$} &
\multicolumn{1}{c}{$n$} &
\multicolumn{1}{c}{$\rho_0$} &
\multicolumn{1}{c}{$\rho_1$} &
\multicolumn{1}{c}{$\rho_2$} \\
\hline
10.0 & 1 &$-0.2972(4)$ & $ 0.315(9)$ & $-3.982(24)$\\
 7.0 & 1 &$-0.3368(3)$ & $ 0.212(6)$ & $-3.902(16)$\\
 4.0 & 1 &$-0.3443(3)$ & $-0.032(6)$ & $-3.525(8)$\\
 4.0 & 2 &$-0.3679(3)$ & $-0.038(6)$ & $-3.251(8)$\\
 3.5 & 2 &$-0.3635(3)$ & $-0.102(6)$ & $-3.109(8)$\\
 3.0 & 2 &$-0.3533(3)$ & $-0.161(6)$ & $-2.937(8)$\\
 2.7 & 2 &$-0.3433(3)$ & $-0.214(6)$ & $-2.813(4)$\\
 2.5 & 2 &$-0.3341(3)$ & $-0.253(6)$ & $-2.717(4)$\\
 2.0 & 2 &$-0.3002(3)$ & $-0.343(4)$ & $-2.423(3)$\\
 1.7 & 2 &$-0.2691(4)$ & $-0.420(4)$ & $-2.208(3)$\\
 1.6 & 2 &$-0.2571(4)$ & $-0.451(6)$ & $-2.128(3)$\\
 1.2 & 3 &$-0.2096(6)$ & $-0.572(6)$ & $-1.664(2)$\\
 1.0 & 4 &$-0.1703(7)$ & $-0.627(6)$ & $-1.410(2)$\\
 0.8 & 5 &$-0.1069(7)$ & $-0.694(6)$ & $-1.144(1)$
\end{tabular}
\end{center}
\end{table}

\begin{table}
\caption[tabthree]{Values for the coefficients $\rho_0$, $\rho_1$,
 and $\rho_2$ using the NRQCD action of Ref.~\protect\cite{fsubBb}.
\label{tab:three}}
\begin{center}
\begin{tabular}{rrrrl}
\multicolumn{1}{c}{$aM_0$} &
\multicolumn{1}{c}{$n$} &
\multicolumn{1}{c}{$\rho_0$} &
\multicolumn{1}{c}{$\rho_1$} &
\multicolumn{1}{c}{$\rho_2$} \\
\hline
20.0 & 1 &$-0.1532(5)$ & $ 0.69(2)\hspace{1.15ex} $ & $-4.85(4) $\\
12.5 & 1 &$-0.2432(4)$ & $ 0.51(1)\hspace{1.15ex} $ & $-4.82(4) $\\
10.0 & 1 &$-0.2772(4)$ & $ 0.45(1)\hspace{1.15ex} $ & $-4.70(3) $\\
 7.0 & 1 &$-0.3174(4)$ & $ 0.322(7)$ & $-4.38(2) $\\
 5.0 & 2 &$-0.3405(4)$ & $ 0.208(7)$ & $-3.89(1) $\\
 4.0 & 1 &$-0.3372(3)$ & $ 0.140(7)$ & $-3.643(8)$\\
 2.7 & 2 &$-0.3375(4)$ & $ 0.027(4)$ & $-2.859(4)$\\
 2.0 & 2 &$-0.3145(4)$ & $-0.037(4)$ & $-2.339(4)$\\
 1.6 & 2 &$-0.2844(4)$ & $-0.054(4)$ & $-1.986(3)$
\end{tabular}
\end{center}
\end{table}

\end{document}